\documentclass[conference,letterpaper]{IEEEtran}
\usepackage{graphicx,epsf,psfrag}
\usepackage{amsmath,amssymb}
\usepackage{amsfonts}
\usepackage{mathrsfs}
\usepackage{etoolbox}
\usepackage[noadjust]{cite}
\usepackage{comment}
\usepackage{multirow}

\usepackage{latexsym}

\newcommand{\iidsim}{\,\stackrel{\text{iid}}{\sim}\,}

\newcommand{\beq}{\begin{equation}}
\newcommand{\eeq}{\end{equation}}
\newcommand{\be}{\begin{equation}}
\newcommand{\ee}{\end{equation}}
\newcommand{\eps}{\epsilon}

\newcommand{\bi}{\begin{itemize}}
\newcommand{\ei}{\end{itemize}}

\newcommand{\calC}{\mathcal{C}}

\newcommand{\calE}{\mathcal{E}}

\newcommand{\calX}{\mathcal{X}}
\newcommand{\calY}{\mathcal{Y}}
\newcommand{\calZ}{\mathcal{Z}}

\newcommand{\bbE}{\mathbb{E}}

\newcommand{\bbP}{\mathbb{P}}

\DeclareMathAlphabet{\mathbsf}{OT1}{cmss}{bx}{n}
\DeclareMathAlphabet{\mathssf}{OT1}{cmss}{m}{sl}% slanted sans serif

\newcommand{\rve}{\mathsf{e}}

\DeclareSymbolFont{bsfletters}{OT1}{cmss}{bx}{n}  
\DeclareSymbolFont{ssfletters}{OT1}{cmss}{m}{n}
\DeclareMathSymbol{\bsfGamma}{0}{bsfletters}{'000}
\DeclareMathSymbol{\ssfGamma}{0}{ssfletters}{'000}
\DeclareMathSymbol{\bsfDelta}{0}{bsfletters}{'001}
\DeclareMathSymbol{\ssfDelta}{0}{ssfletters}{'001}
\DeclareMathSymbol{\bsfTheta}{0}{bsfletters}{'002}
\DeclareMathSymbol{\ssfTheta}{0}{ssfletters}{'002}
\DeclareMathSymbol{\bsfLambda}{0}{bsfletters}{'003}
\DeclareMathSymbol{\ssfLambda}{0}{ssfletters}{'003}
\DeclareMathSymbol{\bsfXi}{0}{bsfletters}{'004}
\DeclareMathSymbol{\ssfXi}{0}{ssfletters}{'004}
\DeclareMathSymbol{\bsfPi}{0}{bsfletters}{'005}
\DeclareMathSymbol{\ssfPi}{0}{ssfletters}{'005}
\DeclareMathSymbol{\bsfSigma}{0}{bsfletters}{'006}
\DeclareMathSymbol{\ssfSigma}{0}{ssfletters}{'006}
\DeclareMathSymbol{\bsfUpsilon}{0}{bsfletters}{'007}
\DeclareMathSymbol{\ssfUpsilon}{0}{ssfletters}{'007}
\DeclareMathSymbol{\bsfPhi}{0}{bsfletters}{'010}
\DeclareMathSymbol{\ssfPhi}{0}{ssfletters}{'010}
\DeclareMathSymbol{\bsfPsi}{0}{bsfletters}{'011}
\DeclareMathSymbol{\ssfPsi}{0}{ssfletters}{'011}
\DeclareMathSymbol{\bsfOmega}{0}{bsfletters}{'012}
\DeclareMathSymbol{\ssfOmega}{0}{ssfletters}{'012}

\newcommand{\hatm}{\hat{m}}

\newcommand{\tilm}{\tilde{m}}

\newcommand{\tilR}{\tilde{R}}

\newcommand{\hatz}{\hat{z}}

\newcommand{\barp}{\bar{p}}

\newcommand{\barV}{\bar{V}}

\DeclareMathOperator{\var}{Var}

\ifcsmacro{theorem}{}{
\newtheorem{theorem}{Theorem}
\newtheorem{lemma}{Lemma}
\newtheorem{proposition}{Proposition}
\newtheorem{corollary}{Corollary}
 
\newtheorem{example}{Example}

}

\newcommand{\qednew}{\nobreak \ifvmode \relax \else
      \ifdim\lastskip<1.5em \hskip-\lastskip
      \hskip1.5em plus0em minus0.5em \fi \nobreak
      \vrule height0.75em width0.5em depth0.25em\fi}

\usepackage{color}
\usepackage{hyperref}

\usepackage{bbm}

\newif\ifarxiv
\arxivtrue

\allowdisplaybreaks

\title{Switched Feedback for the Multiple-Access Channel}

\author{%
  \IEEEauthorblockN{Oliver Kosut}
  \IEEEauthorblockA{School of Electrical, Computer \\and Energy Engineering \\
                    Arizona State University\\
                    Tempe, AZ, USA\\
                    Email: okosut@asu.edu}
    \and
    \IEEEauthorblockN{Michael Langberg}
    \IEEEauthorblockA{Department of Electrical Engineering \\
            University at Buffalo \\
            Buffalo, NY, USA\\
            Email: mikel@buffalo.edu}
    \and
  \IEEEauthorblockN{Michelle Effros}
  \IEEEauthorblockA{Department of Electrical Engineering\\ 
                    California Institute of Technology\\
                    Pasadena, CA, USA\\
                    Email: effros@caltech.edu}
}

\begin{document}

\maketitle

\begin{abstract}
A mechanism called {\em switched feedback} is introduced; under switched feedback, each channel output goes forward to the receiver(s) or {back} to the transmitter(s) but never both. By studying the capacity of the Multiple-Access Channel (MAC) with switched feedback, this work investigates the {benefits of feedback, seeking to maximize} that benefit under reliable and unreliable feedback scenarios. The study is {used to explore} the tradeoffs between cooperation and transmission in the context of communication systems.
Results include upper and lower bounds on the capacity region of the MAC with switched feedback.
\end{abstract}

\section{Introduction}

\begin{figure*}[t]
    \centerline{\includegraphics[width=4.25in]{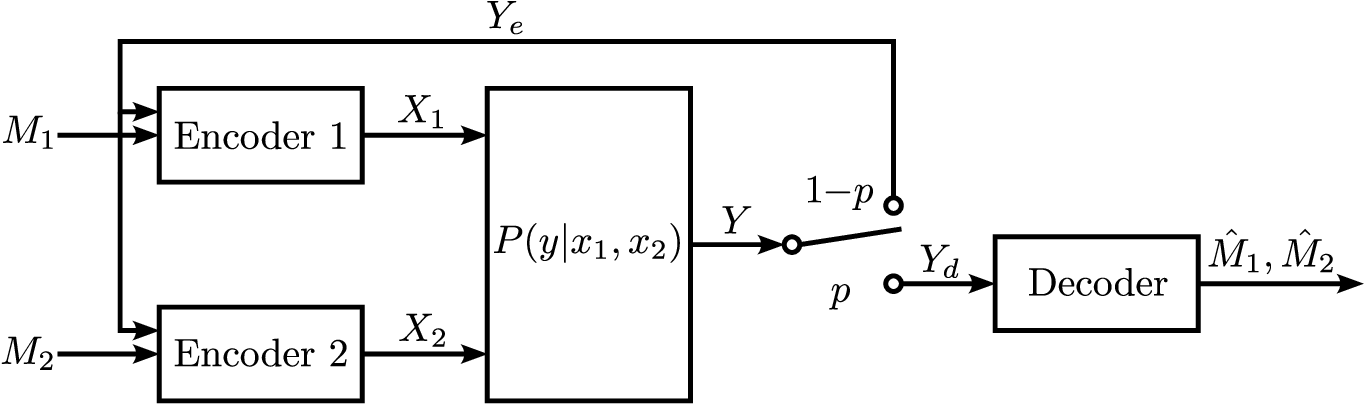}}
    \caption{The switched feedback model for the multiple-access channel. In this example, the feedforward probability $p$ is fixed over time.}
    \label{switched_feedback_diagram}
\end{figure*}

{While most modern communication systems can accommodate feedback, they tend to use it very little and only in primitive and inefficient ways.  This is sometime justified by technological constraints --- computational cost, asymmetries in power resources among communicating devices, or the need to co-exist with legacy technologies --- but our limited understanding of how feedback works, where it is most useful, and how much benefit it can provide may also play a role.  We here propose and study a toy model called ``switched feedback'' designed to help us isolate and study these questions.  

In a multiple-access channel (MAC) with switched feedback (see Fig.~\ref{switched_feedback_diagram})  every channel output goes either to the channel transmitters (feedback) or the channel receiver (feedforward) but not to both.  The system model specifies, for each time step, the feedforward probability.  A few examples follow;  detailed definitions and discussions appear in later sections.

Consider a MAC with switched feedback in which the feedforward probability at each time step is either one (yielding a forward transmission) or zero (yielding feedback).  In this scenario, the feedback times are known, and we can explore how the transmission strategy should differ between feedback and feedforward times steps.  By varying the fraction of time steps spent on forward transmissions, we can explore the relative value of feedforward and feedback communication.  (The MAC model values only the former, but feedback enables cooperation, which can increase the forward capacity.) The insights that arise from these studies may be useful, for example, in networks with half duplex receivers, where every time step spent sending feedback is a time step in which no forward transmission can be received. 

Next, consider a family of systems in which the feedforward probability takes values strictly between zero and one.  As in the deterministic case, we can compare systems with varying total probabilities of feedforward and feedback over time and varying locations of the time steps when feedforward transmissions are more likely.  Comparing the capacities of these systems enables us to consider whether the value of feedback and the best times to take advantage of it differ when the opportunity to engage feedback is uncertain.  They also provide a framework for exploring how the transmitters can balance the potentially competing demands of enabling cooperation and actively cooperating. Such insights may be useful, for example, in networks where feedback to the transmitters may be sporadically but unreliably available, for example due to the potential engagement of communicators in the network that can overhear channel outputs and help by transmitting feedback to the transmitters when otherwise unengaged. }

We study the capacity region $\mathcal{C}$ of switched feedback MACs and present the following main results.
In Section~\ref{sec:problem}, we present a detailed model of the system studied.
In Section~\ref{sec:simple}, we state basic inner and outer bounds to  $\mathcal{C}$ that follow directly from known bounds on the (standard) MAC capacity.
In Section~\ref{sec:binary}, we study the setting in which the  feedforward and feedback times  are governed by an i.i.d. process. Specifically, we present sufficient conditions, over the channel and switching process, that allow us to characterize the MAC capacity region under this model. We also show that these sufficient conditions are satisfied by the binary additive MAC for a certain range of feedforward probabilities. 
In Section~\ref{sec:known}, we study 
the setting in which the  feedforward and feedback times  are governed by a random process that may be time varying but known to the encoders in advance. For any MAC and time varying switching process, we present a characterization of the sum-capacity, alongside inner and outer bounds on the capacity region.
Finally, in Section~\ref{sec:det}, we study the setting in which the switching pattern is deterministic and fixed in advance.
We confirm an intuitive claim in this case, that, roughly speaking, receiving feedback sooner rather than later increases communication rate. \ifarxiv\else In the interest of space, several proofs have been omitted, and are in the extended version \cite{Extended}.\fi

We note that, in our model of study,  the analysis of feedback has strong connections to concepts studied in the context of the two-way channel \cite{shannon1961two}; similarly, in our setting, the analysis of the capacity region is strongly connected to the study of conferencing MACs \cite{willems1983discrete}. We elaborate on these connections in Section~\ref{sec:known}.
The study of cooperation and various forms of feedback has seen significant study in the literature, e.g., \cite{shannon1956zero,dobrushin1958information,gaarder1975capacity,king1978multiple,cover1981achievable, ozarow1984capacity,bross2005improved,venkataramanan2011new,hekstra1989dependence,kramer2006dependence,tandon2009outer,willems1982feedback,kramer1998directed,kramer2003capacity,willems1983discrete,willems1985discrete,noorzad2017unbounded,kosut2023perfect}.
To the best of our knowledge, the models studied in this work have not been addressed in the past.

\section{Problem Description}\label{sec:problem}

\emph{Notation}: For an integer $n$, we use $[n]=\{1,\ldots,n\}$. For a vector $x^n=(x_1,\ldots,x_n)$, we write $x^i=(x_1,\ldots,x_i)$. We use standard notation for entropy and mutual information. The robustly typical set (see \cite{el2011network}) is denoted $T_\eps^{(n)}$; the underlying distribution will be clear from context or specified if not.

A two-transmitter multiple-access channel (MAC) is given by input alphabets $\calX_1,\calX_2$, output alphabet $\calY$, and a conditional probability distribution $P(y|x_1,x_2)$. We assume in this paper that all alphabets are finite. 

In a MAC with switched feedback (see Fig.~\ref{switched_feedback_diagram}), the consequence of the encoders' transmissions at each time $i$ are received at either the transmitters (feedback) or the receiver (feedforward) but never both.  To accommodate switched feedback, we extend the output alphabet from $\calY$ to $\bar\calY=\calY\cup\{\rve\}$, where $\rve$ is an erasure symbol that is distinct from all elements of $\calY$. We use symbol $\rve$ to designate the signal received at the transmitters when the true channel output is received by the decoder and the signal received by the decoder when the true channel output is received by the transmitters. Therefore, the feedback received at the encoders and the channel output received at the decoder are always elements of $\bar\calY$, meaning they can be either elements of $\calY$ or  erasures.

A distinctive element of our model is a \emph{feedforward probability function} $p(t)$, which is a {Reimann-integrable} function from $[0,1]$ to $[0,1]$. The probability function $p(t)$ essentially describes the probability with which the transmitted signal is sent to the decoder (in which case erasure symbols are sent to the encoders) after a fraction $t$ of the blocklength has transpired. We make this precise as follows. We denote by $Y_i$ the raw channel output at time step $i$, which is derived from the channel inputs $X_{1,i},X_{2,i}$ and the channel model $P(y|x_1,x_2)$. The received signal at the decoder is denoted $Y_{d,i}$, and the feedback signal received by both encoders is denoted $Y_{e,i}$. Note that $Y_i\in\calY$ but $Y_{d,i},Y_{e,i}\in\bar\calY$. In a code of blocklength $n$, Bernoulli random variables $V_1,\ldots,V_n$, are drawn independently as $V_i\sim\text{Ber}(p_i)$, where
\begin{equation}\label{switching_probabilities}
    p_i=n\int_{(i-1)/n}^{i/n} p(t)dt
\end{equation}
describes the outcome of the random ``feedforward or feedback'' coin flip at each time $i$.  More precisely, at time step $i$, the received signals at the encoders and decoder are:
\begin{equation}
\begin{array}{lll}
    \text{if }V_i=1\text{, then}& Y_{d,i}=Y_i,& Y_{e,i}=\rve,\\[.5ex]
    \text{if }V_i=0\text{, then}&Y_{d,i}=\rve,& Y_{e,i}=Y_i.
    \end{array}
\end{equation}

An $(n,R_1,R_2)$ code for the two-transmitter MAC with switched feedback consists of $2n$ encoding functions:
\begin{equation}
    x_{j,i}:[2^{nR_j}]\times \bar\calY^{i-1}\to \calX_j,\quad j=1,2,\ i=1,\ldots,n
\end{equation}
and a decoding function
\begin{equation}
    g:\bar\calY^n\to [2^{nR_1}]\times [2^{nR_2}].
\end{equation}
The messages $M_1,M_2$ transmitted by the two encoders are selected independently and uniformly from $[2^{nR_1}]$ and $[2^{nR_2}]$ respectively.
The transmitted signals are derived from messages and feedback as
$
    X_{j,i}=x_{j,i}(M_j, Y_e^{i-1}),
$ for $j=1,2$, $i=1,\ldots,n$.
Message estimates are decoded at the receiver as
$
    (\hat{M}_1,\hat{M}_2)=g(Y_d^n).
$
The probability of error is
$
    P_e=\bbP\{(\hat{M}_1,\hat{M}_2)\ne (M_1,M_2)\}.
$
Rate pair $(R_1,R_2)$ is achievable if there exists a sequence of $(n,R_1,R_2)$ codes with error probability going to $0$ as $n$ grows without bound. The capacity region $\calC$ is the closure of the set of achievable rate pairs. Note that the capacity region depends on the raw MAC model $P(y|x_1,x_2)$ and the feedforward probability function $p(t)$.

\section{Simple Bounds}
\label{sec:simple}

{In this section we present two simple bounds (inner and outer) on the capacity region for the switched-feedback model. We omit the proofs, as they are fairly standard.
We can derive an outer bound by noting that} if we allow the two encoders to fully share information, the problem reduces to a point-to-point erasure channel with erasure probability $1-p_{\text{avg}}$ where
\begin{equation}
    p_{\text{avg}}=\int_{0}^1 p(t)dt.
\end{equation}
This immediately gives the following outer bound.
\begin{proposition}\label{cooperation_outer_bound}
    For any $(R_1,R_2)\in\calC$,
    \begin{equation}\label{eq:cooperation_outer_bound}
        R_1+R_2\le \max_{P(x_1,x_2)} p_{\text{avg}}\, I(X_1,X_2;Y).
    \end{equation}
\end{proposition}

We can also derive the following simple inner bound by ignoring the feedback and using a standard MAC code.
\begin{proposition}
    For any $P(u)P(x_1|u)P(x_2|u)$, the rate pair $(R_1,R_2)\in\calC$ if
    \begin{align}
        R_1+R_2&\le p_{\text{avg}} I(X_1,X_2;Y|U),\\
        R_1&\le p_{\text{avg}} I(X_1;Y|U,X_2),\\
        R_2&\le p_{\text{avg}} I(X_2;Y|U,X_1).
    \end{align}
\end{proposition}

\section{The Capacity Region of a Class of Channels}
\label{sec:binary}

The following theorem gives a class of channels for which the outer bound in Prop.~\ref{cooperation_outer_bound} is achievable.

\begin{theorem}
\label{thm:gen_adder}
Consider a MAC $p(y|x_1,x_2)$ over alphabets $\calX_1$, $\calX_2$, $\calY$ for which, for any distribution $P(x_1,x_2)$ over $\calX_1 \times \calX_2$, it holds that $H(X_1|Y,X_2)=H(X_2|Y,X_1)=0$.
\footnote{We note that the condition $H(X_1|Y,X_2)=H(X_2|Y,X_1)=0$ does not necessarily imply that the given MAC is deterministic.}
Let $p(t)=p$ be a constant feedforward probability function.
Then for any $P(x_1,x_2)$, any 
\begin{align}
R_1 \le \min\{H(X_1|X_2),\ pI(X_1,X_2;Y)\},\label{gen_adder_R1}\\
R_2 \le \min\{H(X_2|X_1),\ pI(X_1,X_2;Y)\},\label{gen_adder_R2}
\end{align}
and any $\alpha \in [0,1]$, $(\alpha R_1,(1-\alpha)R_2) \in \cal C$.
If, in addition, for some
\be
P(x_1,x_2) = \arg\max_{P(x_1,x_2)}I(X_1,X_2;Y)
\ee
it holds that 
\be\label{gen_adder_condition}
pI(X_1,X_2;Y) \leq \max\{H(X_1|X_2),H(X_2|X_1)\},
\ee
then $\calC$ consists of all $(R_1,R_2)$ satisfying
\be\label{gen_adder_R1R2}
R_1+R_2\le \max_{P(x_1,x_2)} pI(X_1,X_2;Y).
\ee
\end{theorem}

\begin{IEEEproof}
The bulk of the proof involves showing that $(R_1,0)\in\calC$ if $R_1$ satisfies \eqref{gen_adder_R1}. Swapping the two indices shows that $(0,R_2)\in\calC$ if $R_2$ satisfies \eqref{gen_adder_R2}.  Then, time sharing can be used to achieve $(\alpha R_1,(1-\alpha)R_2)$ for any $\alpha\in[0,1]$. If \eqref{gen_adder_condition} holds, then from \eqref{gen_adder_R1}--\eqref{gen_adder_R2} one can immediately see that any $(R_1,R_2)$ satisfying \eqref{gen_adder_R1R2} is achievable. The converse follows from Prop.~\ref{cooperation_outer_bound}.

We now show that if $R_1$ satisfies \eqref{gen_adder_R1} then $(R_1,0)\in\calC$. For each time instance $i$, let
\begin{equation}
    Z_i=\begin{cases} (X_{1,i},X_{2,i}) & V_i=0,\\ \rve & V_i=1.\end{cases}
\end{equation}
Let $\calZ=(\calX_1\times\calX_2)\cup\{\rve\}$ be the alphabet for this variable. Note that each encoder can always determine $Z_i$ after time step $i$: if the feedback symbols are erasures $(V_i=1)$, then each encoder knows it so they can determine that $Z_i=\rve$; if the feedback symbols are not erasures, then they equal $Y_i$ and, since by assumption $H(X_{1,i}|Y_i,X_{2,i})=0$, encoder $2$ can determine $X_{1,i}$, and similarly encoder $1$ can determine $X_{2,i}$. Thus, $Z_i$ constitutes shared information at the two encoders.

\emph{Block-Markov structure:} We work over $B$ sub-blocks, each of length $n$. The message for encoder $1$ (the only message in the code, since $R_2=0$) is $m_1=(m_{1,1},\ldots,m_{1,B-1})$ where $m_{1,b}\in[2^{nR_1}]$ for each $b\in[B-1]$. By convention we also let $m_{1,B}=1$. This makes for a total of $(B-1)nR_1$ message bits over a total blocklength $Bn$.

\emph{Codebook generation:} Fix any $P(x_1,x_2)$ and a rate $R_0$ to be determined.  Let $x_2^n(m_0)\iidsim P(x_2)$ for all $m_0\in [2^{nR_0}]$. Let $x_1^n(m_1|m_0)\sim \prod_{i=1}^n P(x_{1,i}|x_{2,i}(m_0))$ for all $m_0\in [2^{nR_0}], m_1\in [2^{nR_1}]$. Let $m_0(z^n)\sim \text{Unif}[2^{nR_0}]$ for all $z^n\in \calZ^n$.

\emph{Encoding:} Prior to block $b$, each encoder knows $z^n(b-1)$, so it can compute $m_{0,b}=m_0(z^n(b-1))$. In the first block, take $m_{0,b}=1$. During block $b$, encoder 1 sends $x_1^n(m_{1,b}|m_{0,b})$, and encoder 2 sends $x_2^n(m_{0,b})$.

\emph{Decoding:} After block $b$ for $b\ge 2$, given received signal $y_d^n(b)$, the decoder finds $\hatm_{0,b}\in[2^{nR_0}]$ as the smallest value such that $(x_2^n(\hatm_{0,b}), y_d^n(b))\in T_\eps^{(n)}$. Then, it finds $\hatm_{1,b-1}\in [2^{nR_1}]$ and $\hatz^n(b-1)\in \calZ^n$ such that 
\begin{gather}
    (x_1^n(\hatm_{1,b-1}|\hatm_{0,b-1}), x_2^n(\hatm_{0,b-1}), \hatz^n(b-1), y_d^n(b-1))\nonumber
    \\ \hspace{2in}\in T_\eps^{(n)},\label{decoding_conditions1}\\
    \hat{m}_{0,b}=m_0(\hatz^n(b-1)).\label{decoding_conditions2}
\end{gather}

In \ifarxiv Appendix~\ref{appendix:gen_adder}\else \cite{Extended}\fi, we complete the proof by showing that the probability of error vanishes as $n\to\infty$ as long as $R_1$ is smaller than the right-hand side of \eqref{gen_adder_R1}.
\end{IEEEproof}

We illustrate Theorem~\ref{thm:gen_adder} with two examples.  In one, the condition for the theorem giving the capacity region, \eqref{gen_adder_condition}, holds for $p$ below a threshold.  In the other, this condition does not hold even for very small $p$.

\begin{example}
Consider the binary additive MAC, given by $\calX_1=\calX_2=\{0,1\}$, $\calY=\{0,1,2\}$, where $Y=X_1+X_2$. Let $(X_1,X_2)$ be a double-symmetric binary source with parameter $q$ (i.e., $P_{X_1,X_2}(x,x)=(1-q)/2,\ P_{X_1,X_2}(x,1-x)=q/2$ for $x=0,1$), then $H(X_1|X_2)=H(q)$ and $H(Y)=H(q)+1-q$, which means, by Theorem~\ref{thm:gen_adder}, we can achieve any $(R_1,R_2)$ if
\begin{equation}
    R_1+R_2\le \max_{q\in[0,1]} \min\{H(q),\ p(H(q)+1-q)\}.
\end{equation}
If we choose $q=1/3$, then $P(x_1,x_2)$ achieves $\max_{P(x_1,x_2)} I(X_1,X_2;Y)=\log 3$. Thus, \eqref{gen_adder_condition} becomes
$
p\log 3\le H(1/3),
$
which corresponds to $p\le 0.5794$. In this case, the capacity region matches the outer bound in Prop.~\ref{cooperation_outer_bound}.
\end{example}

\begin{example}
For any $\epsilon>0$, we present an example MAC for which the first requirement $H(X_1|Y,X_2)=H(X_2|Y,X_1)=0$ holds, but the second requirement does not for any $p>\epsilon$.

Let $\alpha$ be a sufficiently large integer, let $m=\alpha2^\alpha$, and let $\calX_1=\calX_2=\{0,1,\dots,m-1\}$.
Consider the deterministic MAC $p(y|x_1,x_2)$ for which $y=y(x_1,x_2)$ is determined by
\begin{equation}
    y=\begin{cases} (x_1,x_2) & \lfloor x_1/\alpha\rfloor=\lfloor x_2/\alpha\rfloor,\\ 
    (x_1+x_2)_{\hspace{-3mm}\mod m} & \text{otherwise}.\end{cases}
\end{equation}
As $y$ deterministically depends on $x_1$ and $x_2$ it holds for any $P(x_1,x_2)$ that  $H(X_1|Y,X_2)=H(X_2|Y,X_1)=0$.
In addition, $I(X_1,X_2;Y)$
is optimized for any $P(x_1,x_2)$ for which $Y$ is uniform, implying that $I(X_1,X_2;Y)= \log{(\alpha m+m)}=\log{m}+\log{(\alpha+1)}$.

Consider an optimizing $P(x_1,x_2)$.
Let $S$ be the random variable that is $0$ if $Y \in \{0,\dots,m-1\}$ and 1 otherwise.
It holds that $\Pr[S=0]=\frac{1}{\alpha+1}$.
Using $S$, we now analyze $H(X_2|X_1)$.
Specifically,
$H(X_2)\geq H(X_2|S) \geq \Pr[S=1]H(X_2|S=1) = \frac{\alpha\log{m}}{\alpha+1}$,  implying that $H(X_1|X_2)=H(X_1,X_2)-H(X_2)
= 
H(S) + H(X_1,X_2|S) -H(X_2) \leq
1+\frac{\alpha\log{(\alpha m)}}{\alpha+1}+\frac{\log{m^2}}{\alpha+1} -\frac{\alpha\log{m}}{\alpha+1} = 1+\frac{\alpha\log{\alpha}}{\alpha+1}+\frac{2\log{m}}{\alpha+1}$.
Similarly for $H(X_2|X_1)$.
Let $\epsilon>0$.
Taking $\alpha$ to be sufficiently large implies, for any $p>\epsilon$, that the second requirement of Theorem~\ref{thm:gen_adder} is violated.
\end{example}

\section{Known Switching Pattern}
\label{sec:known}

In this section we consider the variant of the problem in which the two encoders know the switching pattern in advance of the coding block. 
{Specifically, $V_i$ are drawn independently as $V_i\sim\text{Ber}(p_i)$, and each $V_i$ is known to the two encoders in advance.}
Let $\calC_{\text{KSP}}$ be the capacity region for this variant, and $C_{\text{KSP,sum}}$ be the corresponding sum-capacity. Since the encoders have more information in this setting, certainly $\calC\subset\calC_{\text{KSP}}$.

The following theorem gives a complete characterization of $C_{\text{KSP,sum}}$ for any feedforward probability function $p(t)$. This theorem shows that this problem is closely related to two others: the two-way channel with a common output, which is used for sharing information between encoders when there is feedback; and the MAC with conferencing, which is related because it also involves exploiting direct transmission between the encoders in order to transmit to the decoder.
Let $\calC_{\text{TW}}$ be the capacity region of the two-way channel given by $P(y|x_1,x_2)$, where $Y$ is used for both outputs of the two-way channel. Also let $C_{\text{TW,sum}}$ be the sum-capacity for this two-way channel. Note that in general, $\calC_{\text{TW}}$ and $C_{\text{TW,sum}}$ are not known; however, we will still express $\calC_{\text{KSP,sum}}$ based on them.

The MAC with conferencing allows the two encoders to transmit directly to each other over links of capacities $C_{12}$ and $C_{21}$ prior to transmitting into the channel. The capacity region for this problem was fully characterized by \cite{willems1983discrete}, and is given by the set of $(R_1,R_2)$ where
\begin{align}
    R_1&\le I(X_1;Y|X_2,U)+C_{12},\\
    R_2&\le I(X_2;Y|X_1,U)+C_{21},\\
    R_1+R_2&\le \min\{I(X_1,X_2;Y|U)+C_{12}+C_{21},\nonumber\\ 
    &\qquad\qquad I(X_1,X_2;Y)\}
\end{align}
for some $P(u)P(x_1|u)P(x_2|u)$. The main difference between the conferencing model and ours is that with conferencing, communication between encoders happens entirely before communication to the decoder, whereas in our model these two aspects are intertwined. Thus, the capacity region of the conferencing model does not directly lead to the capacity for the switching model, but one can see remnants of the characterization of the conferencing capacity in our result.

\begin{theorem}\label{thm:ksp}
    The sum-capacity for the known switching pattern model with feedback probability function $p(t)$ is 
    \begin{multline}
        C_{\text{KSP,sum}}
        =\max \min_{\tau\in [0,1]} \int_0^\tau \Big[(1-p(t))C_{\text{TW,sum}}
        \\+p(t) I(X_1(t),X_2(t);Y(t)|U(t))\Big]dt
        \\+\int_{\tau}^1 p(t) I(X_1(t),X_2(t);Y(t))dt
    \end{multline}
    where the max is over all distributions 
    \be\label{distributions}
    P(u(t))\,P(x_1(t)|u(t))\,P(x_2(t)|u(t))
    \ee
    for each $t\in[0,1]$.
\end{theorem}

To prove this theorem, we will give an achievability lemma and a converse lemma, proved in \ifarxiv Appendices~\ref{appendix:achievability} and~\ref{appendix:converse} respectively\else \cite{Extended}\fi. We will then show that the bounds from the two lemmas coincide in the sum-rate.

\begin{lemma}\label{lemma:achievability}
    Consider any set of distributions given by \eqref{distributions} for all $t\in[0,1]$. Then $(R_1,R_2)\in\calC_{\text{KSP}}$ if for some $(S_1,S_2)\in\calC_{\text{TW}}$,
    \begin{align}
        &R_1+R_2\le \min_{\tau\in [0,1]} \int_0^\tau \Big[(1-p(t))(S_1+S_2)\nonumber
        \\&\quad\quad\quad+p(t) I(X_1(t),X_2(t);Y(t)|U(t))\Big]dt\nonumber
        \\&\quad\quad\quad+\int_{\tau}^1 p(t) I(X_1(t),X_2(t);Y(t))dt,\label{achievability_integral_R1R2}\\
        &R_1\le \int_0^1 \big[(1-p(t))S_1+p(t) I(X_1(t);Y(t)|U(t),X_2(t))\big]dt,\label{achievability_integral_R1}\\
        &R_2\le \int_0^1 \big[(1-p(t))S_2+p(t) I(X_2(t);Y(t)|U(t),X_1(t))\big]dt.\label{achievability_integral_R2}
    \end{align}
\end{lemma}

\begin{lemma}\label{lemma:converse}
    If $(R_1,R_2)\in\calC_{\text{KSP}}$, then there exists distributions as in \eqref{distributions} where for all $t\in[0,1]$, there exists $(S_1(t),S_2(t))\in\calC_{\text{TW}}$ such that 
        \begin{align}
        &R_1+R_2\le \min_{\tau\in [0,1]} \int_0^\tau \Big[(1-p(t))(S_1(\tau)+S_2(\tau))\nonumber
        \\&\qquad+p(t) I(X_1(t),X_2(t);Y(t)|U(t))\Big]dt\nonumber
        \\&\qquad+\int_{\tau}^1 p(t) I(X_1(t),X_2(t);Y(t))dt,\label{converse_R1R2}\\
        &R_1\le \int_0^1 \big[(1-p(t))S_1(1)\nonumber
        \\&\qquad+p(t) I(X_1(t);Y(t)|U(t),X_2(t))\big]dt,\label{converse_R1}\\
        &R_2\le \int_0^1 \big[(1-p(t))S_2(1)\nonumber
        \\&\qquad+p(t) I(X_2(t);Y(t)|U(t),X_1(t))\big]dt.\label{converse_R2}
    \end{align}
\end{lemma}

\begin{IEEEproof}[Proof of Theorem~\ref{thm:ksp}]
The upper bound on the sum-capacity follows immediately from \eqref{converse_R1R2}, since $S_1(t)+S_2(t)\le C_{\text{TW,sum}}$ for any $(S_1(t),S_2(t))\in\calC_{\text{TW}}$. To show that Lemma~\ref{lemma:achievability} gives the same sum-rate, note that the sum of the right-hand sides of \eqref{achievability_integral_R1} and \eqref{achievability_integral_R2} is
\begin{align}
 &\int_0^1 \big[(1-p(t))(S_1+S_2)+p(t) \big(I(X_1(t);Y(t)|U(t),X_2(t))\nonumber
 \\&\qquad+p(t) I(X_2(t);Y(t)|U(t),X_1(t))\big)\big]dt
 \\&\ge \int_0^1 \big[(1-p(t))(S_1+S_2)\nonumber
 \\&\qquad+p(t) I(X_1(t),X_2(t);Y(t)|U(t))\big] dt,
\end{align}
where the inequality follows from the fact that $X_{1}(t),X_{2}(t)$ are conditionally independent given $U(t)$. This means that for any sum-rate satisfying \eqref{achievability_integral_R1R2} for $\tau=1$, there exist $R_1,R_2$ achieving that sum-rate that also satisfy \eqref{achievability_integral_R1}--\eqref{achievability_integral_R2}. Taking $(S_1,S_2)\in\calC_{\text{TW}}$ where $S_1+S_2=C_{\text{TW,sum}}$ completes the proof.
\end{IEEEproof}

\vspace{-.1in}
\section{Deterministic switching pattern}
\label{sec:det}

The known switching pattern model studied in the previous section is closely related to the case where the switching pattern is deterministic; i.e., the feedforward probability function $p(t)$ takes values only in $\{0,1\}$. This relationship is captured in the following proposition.

\begin{proposition}
    Let $p(t)$ be a feedforward probability function taking values in $\{0,1\}$ with a finite number of points of discontinuity. Then $\calC=\calC_{\text{KSP}}$.
\end{proposition}

\begin{IEEEproof}
Recall the model in \eqref{switching_probabilities} that, in a block of length $n$, the switching probability at timestep $i$ is the average of $p(t)$ over the corresponding interval of length $1/n$. Thus, assuming $p(t)\in\{0,1\}$, and $p(t)$ has $k$ points of discontinuity, then $p_i\in\{0,1\}$ for at least $n-k$ timesteps $i$. If we restrict coding to these timesteps, then the encoders can predict the switching pattern in advance, since the $V_i$ variables are deterministic. Taking $n$ large enough, the $k$ unused timesteps are irrelevant to achievable rates. Thus $\calC_{\text{KSP}}\subset\calC$. Since the converse is trivially true, we have proved that $\calC=\calC_{\text{KSP}}$.
\end{IEEEproof}

We can use Lemmas~\ref{lemma:achievability} and~\ref{lemma:converse} to derive the following result, which shows that, given $p_{\text{avg}}$, the best switching pattern is the one where all the feedback precedes all the feedforward.

\begin{corollary}
    For any feedforward probability function $p(t)$, if $(R_1,R_2)\in\calC$, then there exist $(S_1,S_2)\in\calC_{\text{TW}}$ where
    \begin{align}
        R_1+R_2&\le (1-p_{\text{avg}})(S_1+S_2)+p_{\text{avg}}I(X_1,X_2;Y|U),\label{cor1}\\
        R_1+R_2&\le p_{\text{avg}} I(X_1,X_2;Y),\label{cor2}\\
        R_1&\le (1-p_{\text{avg}}) S_1+p_{\text{avg}}I(X_1;Y|U,X_2),\label{cor3}\\
        R_2&\le (1-p_{\text{avg}}) S_2+p_{\text{avg}}I(X_2;Y|U,X_1),\label{cor4}
    \end{align}
    for some $P(u)P(x_1|u)P(x_2|u)$. Moreover, the region described above is precisely $\calC$ if the feedforward probability function is given by
    \be\label{step_switching_pattern}
        p(t)=\begin{cases} 0, & t<1-p_{\text{avg}}, \\ 1, & t\ge 1-p_{\text{avg}}.\end{cases}
    \ee
\end{corollary}
\begin{IEEEproof}
Since $\calC\subset\calC_{\text{KSP}}$, we can apply the outer bound in Lemma~\ref{lemma:converse}. In particular, there exist variables $(U(t),X_1(t),X_2(t),Y(t))$ and $(S_1(t),S_2(t))\in\calC_{\text{TW}}$ satisfying \eqref{converse_R1R2}--\eqref{converse_R2}. Let $T\sim\text{Unif}(0,1)$, $U=(U(T),T)$, $X_1=X_1(T)$, $X_2=X_2(T)$, $Y=Y(T)$. Also let $Y_d(t),Y_e(t)$ be the output of the switch with input $Y(t)$ (i.e., each is either $Y(t)$ or $\rve$) and feedforward probability $p(t)$. Then let $Y_d=Y_d(T)$. Now we can write
\begin{align}
    &\int_0^1 p(t) I(X_1(t),X_2(t);Y(t)|U(t))dt
    \\&=\int_0^1 I(X_1(t),X_2(t);Y_d(t)|U(t))dt
    \\&=I(X_1(T),X_2(T);Y_d(T)|U(T))
    \\&=I(X_1,X_2;Y_d|U)
    \\&=p_{\text{avg}} I(X_1,X_2;Y|U)
\end{align}
where the last step follows since $p_{\text{avg}}=\bbP(Y_d(T)=\rve)$. Thus setting $\tau=1$ in \eqref{converse_R1R2} we get
\be
R_1+R_2\le (1-p_{\text{avg}})(S_1(1)+S_2(1))+p_{\text{avg}}I(X_1,X_2;Y|U).
\ee
Similarly, setting $\tau=0$ in \eqref{converse_R1R2} gives
\begin{align}
R_1+R_2&\le I(X_1,X_2;Y_d|T)
\le I(U,X_1,X_2;Y_d)
\\&=I(X_1,X_2;Y_d)
=p_{\text{avg}} I(X_1,X_2;Y).
\end{align}
Finally, \eqref{converse_R1}--\eqref{converse_R2} become
\begin{align}
    R_1&\le (1-p_{\text{avg}})S_1(1)+p_{\text{avg}}I(X_1;Y|U,X_2),\\
    R_2&\le (1-p_{\text{avg}})S_2(1)+p_{\text{avg}}I(X_2;Y|U,X_1).
\end{align}

To show that this same region is achievable with $p(t)$ given by \eqref{step_switching_pattern}, we apply the inner bound in Lemma~\ref{lemma:achievability}. Let $(U(t),X_1(t),X_2(t))$ have a distribution independent of $t$. The sum-rate condition in \eqref{achievability_integral_R1R2} becomes, for $\tau\le 1-p_{\text{avg}}$,
\begin{align}\label{tau_small}
    R_1+R_2\le \tau(S_1+S_2)+p_{\text{avg}}I(X_1,X_2;Y)
\end{align}
and for $\tau\ge 1-p_{\text{avg}}$,
\begin{multline}\label{tau_large}
    R_1+R_2\le (1-p_{\text{avg}})(S_1+S_2)+(\tau-1+p_{\text{avg}})I(X_1,X_2;Y|U)
    \\+(1-\tau) I(X_1,X_2;Y).
\end{multline}
Note that \eqref{tau_small} is increasing in $\tau$, whereas \eqref{tau_large} is decreasing in $\tau$ (since $I(X_1,X_2;Y|U)\le I(X_1,X_2;Y)$), so the two extremes $\tau=0,1$ are the strictest conditions. The $\tau=1$ condition becomes \eqref{cor1}, and the $\tau=0$ condition becomes \eqref{cor2}. Furthermore, \eqref{achievability_integral_R1} and \eqref{achievability_integral_R2} become \eqref{cor3} and \eqref{cor4} respectively.
\end{IEEEproof}

\section{Conclusions}
We study the relative value of feedback and feedforward transmissions in the multiple-access channel.  Theorem~\ref{thm:gen_adder} demonstrates sufficient conditions under which we can get the maximal cooperation sum capacity associated with the expected fraction of the blocklength where feedback makes cooperation possible.
Theorem~\ref{thm:ksp} and the proposition that follows it establish the sum-capacity when the switching pattern is known in advance (either for deterministic or stochastic feedback timing). Roughly speaking, the solution these results reach suggests that the capacity is impacted in the first part of the block by the rate at which the two transmitters can cooperate while transmitting to the decoder using a non-cooperative code, whereas in the remainder of the block by their ability to transmit to the decoder by taking advantage of previous cooperation.

\section*{Acknowledgments}

This work is supported in part by NSF grants CCF-2107526, CCF-2245204, and by a grant from the Caltech Center for Sensing to Intelligence.

\bibliographystyle{IEEEtran}
\bibliography{MAC_feedback}

\ifarxiv

\clearpage

\appendices

\section{Probability of Error Analysis for the Proof of Theorem~\ref{thm:gen_adder}}\label{appendix:gen_adder}

Consider the decoding process in block $b$. By the packing lemma \cite{el2011network}, the decoder's first step of finding $\hatm_{0,b}$ has small probability of error if
\begin{equation}
    R_0<I(X_2;Y_d).
\end{equation}
In the second step, there are several possible error events. First, the correct $m_{1,b-1}$ and $z^n(b-1)$ may not satisfy the conditions in \eqref{decoding_conditions1}--\eqref{decoding_conditions2}. This occurs with small probability of error due to the law of large numbers. Second, some 
pair $(\hatm_{1,b-1},\hatz^n(b-1))$ may satisfy \eqref{decoding_conditions1}--\eqref{decoding_conditions2} where $\hatm_{1,b-1}\ne m_{1,b-1}$ but $\hatz^n(b-1)=z^n(b-1)$. In this case, \eqref{decoding_conditions2} will necessarily be satisfied, but the probability of satisfying \eqref{decoding_conditions1} for any given $\hatm_{1,b-1}$ is at most
\begin{equation}
    2^{-n(I(X_1;Y_d,Z|X_2)-\eps)}.
\end{equation}
Thus, by the union bound this probability vanishes as long as
\begin{align}\label{first_R1_condition}
    R_1&<I(X_1;Y_d,Z|X_2)
    \\&=H(X_1|X_2)-p H(X_1|X_2,Y_d,Z,V=1)\nonumber
    \\&\qquad-(1-p) H(X_1|X_2,Z,Y_d,V=0)
    \\&=H(X_1|X_2)-p H(X_1|Y,Z,X_2,V=1)\nonumber
    \\&\qquad-(1-p) H(X_1|X_2,Z,Y_d,X_1,V_i=0)\label{x1_given_x2}
    \\&=H(X_1|X_2)\label{x1_given_x2_2}
\end{align}
where \eqref{x1_given_x2} follows because when $V=1$, $Y_{d}=Y$, and when $V=0$, $Z$ includes $X_1$, and \eqref{x1_given_x2_2} follows by the assumption that $H(X_1|Y,X_2)=0$. The final error event is that a pair  $(\hatm_{1,b-1},\hatz^n(b-1))$ satisfy \eqref{decoding_conditions1}--\eqref{decoding_conditions2} where both parts are incorrect. If $\hatz^n(b-1)$ is not itself typical, then \eqref{decoding_conditions1} will never be satisfied. If $\hatz^n(b-1)$ is typical, then the probability of satisfying both conditions is at most
\begin{equation}
    2^{-n(I(X_2,Y_d;Z)+I(X_1;Y_d,Z|X_2)+R_0-\eps)}.
\end{equation}
Since there are at most $2^{n(H(Z)+\eps)}$ typical $\hatz^n(b-1)$, by the union bound the total probability vanishes if
\begin{equation}
    R_1+H(Z) < I(X_2,Y_d;Z)+I(X_1;Y_d,Z|X_2)+R_0.
\end{equation}
Recalling that any $R_0<I(X_2;Y_d)$ is acceptable, we can achieve any $R_1$ as long as we have \eqref{first_R1_condition} and
\begin{align}
    R_1&<I(X_2,Y_d;Z)+I(X_1;Y_d,Z|X_2)+I(X_2;Y_d)-H(Z)
    \\&=-H(Z|X_2,Y_d)+I(X_1;Y_d|X_2)\nonumber\\
    & \ \ \ \ \ \ \ \ \ \ \ \ +I(X_1;Z|X_2,Y_d)+I(X_2;Y_d)
    \\&=I(X_1,X_2;Y_d)-H(Z|X_1,X_2,Y_d)
    \\&=I(X_1,X_2;Y_d)\label{step3}
    \\&=p I(X_1,X_2;Y)
\end{align}
where \eqref{step3} follows since $Z$ is a function of $(X_1,X_2)$ and $V$, and $V$ is a function of $Y_d$. Therefore, we can achieve $(R_1,0)$ if
\begin{equation}
    R_1<\min\{H(X_1|X_2),\ pI(X_1,X_2;Y)\}.
\end{equation}

\section{Proof of Lemma~\ref{lemma:achievability}}\label{appendix:achievability}

We will construct a code of blocklength $nB$ for an integers $B,n$. This overall length is divided into $B$ blocks of length $n$ each. Let
\begin{equation}
    \barp_b=B \int_{(b-1)/B}^{b/B} p(t)dt,\quad b=1,\ldots,B.
\end{equation}
Fix $\epsilon>0$. Let $(S_1,S_2)$ be in the interior of $\calC_{\text{TW}}$. Thus, there exists some $n_0$ such that, for all $n\ge n_0$, there is a two-way channel code of length $n$, rates $(S_1,S_2)$, and probability of error at most $\eps/B$. Moreover, for sufficiently large $n$, for each $b\in[B]$ where $\barp_b<1$, $n(1-\eps)(1-\barp_b)\ge n_0$. Also choose a distribution $P(u_b)P(x_{1,b}|u_b) P(x_{2,b}|u_b)$ for each $b$.

We will show that any rate-pair $(R_1,R_2)$ is achievable if 
\begin{align}
&R_1+R_2<\min_{b_0\in\{1,\ldots,B+1\}}\frac{1}{B}\sum_{b'=1}^{{b_0}-1} \big[(1-\eps)(1-\barp_{b'})(S_1+S_2)\nonumber
\\&\qquad+\barp_{b'} I(X_{1,b'},X_{2,b'};Y_{b'}|U_{b'})\big]\nonumber
\\&\qquad+\frac{1}{B}\sum_{b'=b_0}^B \barp_{b'} I(X_{1,b'},X_{2,b'};Y_{b'}),\label{achievability_R1R2}\\
&R_1<\frac{1}{B} \sum_{b=1}^B ((1-\eps)(1-\barp_b)S_1+\barp_b I(X_{1,b};Y_b|U_b,X_{2,b})),\label{achievability_R1}\\
&R_2<\frac{1}{B} \sum_{b=1}^B ((1-\eps)(1-\barp_b)S_2+\barp_b I(X_{2,b};Y_b|U_b,X_{1,b}))\label{achievability_R2}
\end{align}
where $Y_b$ is the output of the raw channel $P(y|x_1,x_2)$ with $(X_{1,b},X_{2,b})$ as the inputs. Roughly speaking, this is enough to prove achievability in the lemma, since we can take $\eps$ arbitrarily close to $0$, $S_1+S_2$ arbitrarily close to the boundary of $\calC_{\text{TW}}$, and $B\to\infty$. In particular, by taking $B$ large, the switching probability function $p(t)$ can be well-approximated, and then the summations become integrals.

For $j=1,2$ let
\be
\tilR_{j}=\max\left\{0,B\,R_j-\sum_{b=1}^B (1-\eps)(1-\barp_b)S_{j}\right\}.
\ee
Let
\be
\phi_j:[2^{nBR_j}]\to [2^{n\tilR_j}]\times  \prod_{b=1}^B [2^{n(1-\eps)(1-\barp_b)S_j}] 
\ee
be an injective function; this is possible since the range has at least as many elements as the domain. Given message $m_j\in[2^{nBR_j}]$, let $(\tilm_j,m_{j,1},\ldots,m_{j,B})=\phi_j(m_j)$ where $\tilm_{j}\in [2^{n\tilR_{j}}]$ and $m_{j,b}\in [2^{n(1-\eps)(1-\barp_b)S_{j}}]$. In the coding scheme, $(m_{1,b},m_{2,b})$ will be transmitted using the two-way channel code in block $b$; the remaining part of the messages (if any) not contained in these are represented by $\tilm_1,\tilm_2$. Let
\be
m_{0,b}=(m_{j,b'}:j\in\{1,2\}, b'\in\{1,\ldots,b-1\}).
\ee
Note that $m_{0,b}\in [2^{nR_{0,b}}]$, where
\be
R_{0,b}=\sum_{b'=1}^{b-1} (1-\eps)(1-\barp_b)(S_1+S_2).
\ee
This message $m_{0,b}$ consists of the common knowledge at both encoders prior to block $b$.

\emph{Codebook generation:} For each $b$, select the following codebooks:
\begin{align}
    &u_b^n(m_{0,b})\iidsim p(u_b),\text{ for all } m_{0,b}\in [2^{nR_{0,b}}],\\
    &x_{j,b}^n(m_{j}|m_{0,b})\sim \prod_{i=1}^n p(x_{j,b,i}|u_{b,i}(m_{0,b})),\nonumber\\
    &\qquad \text{for all }m_j\in [2^{nBR_j}], m_{0,b}\in [2^{nR_{0,b}}].
\end{align}

\emph{Encoding}: Each of the $B$ $n$-length blocks consist of a feedback sub-block and a feedforward sub-block, which the encoders know in advance. In the feedback sub-block, the encoders exploit a two-way channel of the proper length to send $(m_{1,b},m_{2,b})$. This works as long as the feedback sub-block includes at least $(1-\eps)(1-\barp_b)n$ timesteps. If it includes less than this, declare an error.

For the feedforward sub-block, from the previous blocks each encoder has $m_{0,b}$ which consists of $(m_{1,b'},m_{2,b'})$ for all $b'<b$. (Assume for now that these have been decoded correctly by the two-way channel codes.) 
Now encoder $j$ transmits
$x_{j,b,i}(m_j|m_{0,b})$
for all time-steps $i$ that involve feed-forward.

\emph{Decoding:} After all $B$ blocks, the decoder has $y^n_d(b)$ for $b\in[B]$. The decoder finds a unique pair $(\hatm_1,\hatm_2)$ such that, 
\begin{multline}\label{decoding_condition}
(u_b^n(\hatm_{0,b}),x_{1,b}^n(\hatm_1|\hatm_{0,b}),x_{2,b}^n(\hatm_2|\hatm_{0,b}),y^n_d(b))\in T_\eps^{(n)}(b) \\\text{for all }b\in[B]
\end{multline}
where the typical set is with respect to the distribution 
\be
P(u_b)P(x_{1,b}|u_b)P(x_{2,b}|u_b)P(y|x_{1,b},x_{2,b})P(v_b)P(y_d(b)|y,v_b)
\ee
where $P(v_b)$ is $\text{Ber}(\barp_b)$, and $P(y_d(b)|y,v_b)$ is the deterministic channel in which $y_d(b)=y$ is if $v_b=1$ and $y_d(b)=\rve$ if $v_b=0$. Note that in the above condition, $\hatm_{0,b}$ is a function of $(\hatm_1,\hatm_2)$ in the same way that $m_{0,b}$ is a function of $(m_1,m_2)$. If there is no pair $(\hatm_1,\hatm_2)$ satisfying the above condition or more than one, declare an error.

\emph{Probability of Error Analysis:} In block $b$, in the $i$th timestep, the feedforward probability is
\be
p_{b,i}= nB \int_{(b-1)/B+(i-1)/(nB)}^{(b-1)/B+i/(nB)} p(t)dt.
\ee
Thus
\be
\barp_b=\frac{1}{n} \sum_{i=1}^n p_{b,i}.
\ee
Denote by $V_{b,i}$ the variable indicating the switching decision at timestep $i$ of block $b$, so $V_{b,i}\sim\text{Ber}(p_{b,i})$. Also let $\barV_b=
\frac{1}{n} \sum_{i=1}^n V_{b,i}$.
Thus $\bbE[\barV_b]=\barp_b$.

Recall that the two messages are broken down into $(m_{1,b},m_{2,b})$ for $b=1,\ldots,B$, as well as $(\tilm_1,\tilm_2)$. The former parts are transmitting using the two-way channel code and so constitute the common knowledge prior to block $b$ contained in $m_{0,b}$. Thus, we can break down possible error events first by whether the decoder makes an error on $m_{0,b}$ for some $b$. Since these common knowledge messages are nested (i.e., $m_{0,b}$ includes $m_{0,b'}$ if $b>b'$), we can divide these events based on the smallest $b$ in which an error occurs on $m_{0,b}$. If the decoder successfully decodes $m_{0,b}$ for all $b$, it may still make an error if it decodes $(\tilm_1,\tilm_2)$ incorrectly. With this in mind, we list the following error events:
\begin{align}
    \calE_{1,b}&=\{V_b^n\notin T_\eps^{(n)}(V_b)\},\\
    \calE_{2,b}&=\{\text{two-way channel code makes an error in block }b\},\\
    \calE_{3}&=\{\text{\eqref{decoding_condition} does not hold with }\hatm_1=m_1,\hatm_2=m_2\}\\
    \calE_{4,b_0}&=\{\text{\eqref{decoding_condition} holds with }\hatm_{0,b'}=m_{0,b'}\text{ for all }b'\le b_0,\nonumber\\
    &\qquad \hatm_{0,b_0+1}\ne m_{0,b_0+1}\},\\
    \calE_5&=\{\text{\eqref{decoding_condition} holds with }\hatm_{0,b'}=m_{0,b'}\text{ for all }b'\in[B+1],\nonumber\\
    &\qquad  \hatm_{1}\ne m_1,\hatm_{2}\ne m_2\},\\
    \calE_6&=\{\text{\eqref{decoding_condition} holds with }\hatm_{0,b'}=m_{0,b'}\text{ for all }b'\in[B],\nonumber\\
    &\qquad  \hatm_{1}\ne m_1,\hatm_{2}= m_2\},\\
    \calE_7&=\{\text{\eqref{decoding_condition} holds with }\hatm_{0,b'}=m_{0,b'}\text{ for all }b'\in[B],\nonumber\\
    &\qquad  \hatm_{1}= m_1,\hatm_{2}\ne m_2\}.
\end{align}

To analyze $\bbP(\calE_{1,b})$, note that $V_b^n\in T_\eps^{(n)}(V_b)$ iff $|\barV_n-\barp_n|\le \eps\max\{\barp_b,1-\barp_b\}$. Thus, by Chebyshev's inequality,
\begin{align}
    \bbP(\calE_{1,b})
    &= \bbP\big(|\barV_b-\barp_b|>\eps\max\{\barp_b,1-\barp_b\}\big)
    \\&\le \frac{\var[\barV_b]}{\eps^2\max\{\barp_b,1-\barp_b\}^2}
    \\&=\frac{\sum_{i=1}^n p_{b,i}(1-p_{b,i})}{ \eps^2 \max\{\barp_b,1-\barp_b\}^2n^2}
    \\&\le \frac{1}{4 \eps^2 \max\{\barp_b,1-\barp_b\}^2n}.
\end{align}
In the case that $\barp_b=1$ or $\barp_b=0$, then $\barV_b$ is deterministic, which means $P(\calE_{1,b})=0$. Thus, for sufficiently large $n$, $\bbP(\bigcup_b\calE_{1,b})\le \eps$.

If $\calE_{1,b}$ does not occur, then the $b$th feedback sub-block has length at least $(1-\eps)(1-\barp_b)n$, which as argued above, for sufficiently large $n$ is at least $n_0$, which means by our construction of the two-way channel codes, $\bbP(\calE_{2,b}|\calE_{1,b}^c)\le \eps/B$ for each $b$. Thus,
\be
\bbP\left(\bigcup_b \calE_{1,b}\cup \calE_{2,b}\right)\le 2\eps.
\ee
This shows that with probability at least $1-2\eps$, the two-way channel codes are decoded without error, which means that prior to block $b$, each encoder can successfully decode $m_{0,b}$. We assume this holds for all subsequent calculations.

Form our analysis of $\calE_{1,b}$, we have established that with high probability $V_b^n$ is typical. The other variables within the $b$th block are drawn independently from the same distribution given $V_b^n$, so by the law of large numbers, $\bbP(\calE_{3})\le \eps$ for sufficiently large $n$.

The remaining error events are handled with the packing lemma. Consider event $\calE_{4,b_0}$. Consider some $\hatm_1,\hatm_2$ such that $\hatm_{0,b'}=m_{0,b'}$ for all $b'\le b_0$, and $\hatm_{0,b_0+1}\ne m_{0,b_0+1}$. By the definition of $m_{0,b}$, this means that $\hatm_{j,b}=m_{j,b}$ for $j=1,2$ and $b=1,\ldots,b_0-1$, but $(\hatm_{1,b_0},\hatm_{2,b_0})\ne (m_{1,b_0},m_{2,b_0})$. Thus, $u^n(\hatm_{0,b'})$ will be independent from $u^n(m_{0,b'})$ for all blocks $b'\ge b_0$. In addition, $(\hatm_1,\hatm_2)\ne (m_1,m_2)$, so $x_{1,b}^n(\hatm_1|\hatm_{0,b}),x_{2,b}^n(\hatm_1|\hatm_{0,b}$ will be conditionally independent given $u_b^n(\hatm_{0,b})$ for $b<b_0$, and unconditionally independent for $b\ge b_0$. The possible incorrect codewords corresponds to the common information starting from block $b_0$ (i.e., $(m_{1,b'},m_{2,b'})$ for $b'\ge b_0$) and the remaining parts of the messages $\tilm_1,\tilm_2$. The total rate of these incorrect codebook is thus $\sum_{b'=b_0}^B (1-\eps)(1-\barp_b)(S_1+S_2)+\tilR_1+\tilR_2$. On the other side, we can condition on $U_{b'}$ for $b'<b_0$ (since those codewords were decoded correctly), but for $b'\ge b_0$, we cannot condition on this. Thus, by the packing lemma, $\bbP(\calE_{4,b_0})$ vanishes if
\begin{multline}\label{E4_condition}
\sum_{b'=b_0}^B (1-\eps)(1-\barp_b)(S_1+S_2)+\tilR_1+\tilR_2
\\<\sum_{b'=1}^{b_0-1} I(X_{1,b'},X_{2,b'};Y_{d,b'}|U_{b'})
\\+\sum_{b'=b_0}^B I(U_{b'},X_{1,b'},X_{2,b'};Y_{d,b'})
\end{multline}
where $Y_{d,b}$ is the decoder's channel output with $(X_{1,b},X_{2,b})$ as the inputs where $V_b\sim\text{Ber}(\barp_b)$. 
Using similar reasoning, $\bbP(\calE_5)$ vanishes if
\be
\tilR_1+\tilR_2<\sum_{b=1}^B  I(X_{1,b},X_{2,b};Y_{d,b}|U_b).
\ee
We can see this condition as a special case of \eqref{E4_condition} with $b_0=B+1$, which means we need \eqref{E4_condition} to hold for all $b_0\in\{1,\ldots,B+1\}$. Using the definitions of $\tilR_1,\tilR_2$, \eqref{E4_condition} is equivalent to
\begin{multline}
\max\left\{B(R_1+R_2), \sum_{b'=1}^B (1-\eps)(1-\barp_{b'})(S_1+S_2)\right\}
\\<\sum_{b'=1}^{b_0-1} ((1-\eps)(1-\barp_{b'})(S_1+S_2)+\barp_{b'} I(X_{1,b'},X_{2,b'};Y_{b'}|U_{b'}))
\\+\sum_{b'=b_0}^B \barp_{b'} I(X_{1,b'},X_{2,b'};Y_{b'}).
\end{multline}
Thus, by the condition on $R_1+R_2$ in \eqref{achievability_R1R2}, $\bbP(\calE_{4,b_0})$ and $\bbP(\calE_{5})$ vanish.

We have that $\bbP(\calE_6)$ vanishes if
\be
\tilR_1<\sum_{b=1}^B I(X_{1,b};Y_{d,b}|U_b,X_{2,b}).
\ee
Thus, the condition on $R_1$ in \eqref{achievability_R1} implies that $\bbP(\calE_6)$ vanishes. Similarly the condition on $R_2$ in \eqref{achievability_R2} implies that $\bbP(\calE_7)$ vanishes.

\section{Proof of Lemma~\ref{lemma:converse}}\label{appendix:converse}

Consider a code with rate pair $(R_1,R_2)$, blocklength $n$, and probability of error $\eps$. Let $U_i=Y_e^{i-1}$. Since $X_{j,i}$ depends only on $M_j$ and $Y_e^{i-1}$, $X_{1,i}$ and $X_{2,i}$ are conditionally independent given $U_i$. We may express the capacity region of the two-way channel $\calC_{\text{TW}}$ as the set of rate-pairs $(S_1,S_2)$ where, for some $n$, and some distribution 
\begin{multline}
p(m_1)p(m_2)\\ \cdot \prod_{i=1}^n p(x_{1,i}|m_1,y^{i-1}) p(x_{2,i}|m_2,y^{i-1}) p(y_i|x_{1,i},x_{2,i}),
\end{multline}
we have
\begin{align}
    n S_1&\le I(M_1;Y^n|M_2),\\
    n S_2&\le I(M_2;Y^n|M_1).
\end{align}

For our MAC code, 
for each $k\in[n]$, let
\begin{align}
    S_1^k&=\frac{I(M_1;Y_e^k|M_2)}{\sum_{i=1}^k (1-p_i)},\\
    S_2^k&=\frac{I(M_2;Y_e^k|M_1)}{\sum_{i=1}^k (1-p_i)}.
\end{align}
Also, for any $v^k\in\{0,1\}^k$, let
\begin{align}
    S_1^k(v^k)&=\frac{I(M_1;Y_e^k|M_2,V^k=v^k)}{\sum_{i=1}^k (1-p_i)},\\
    S_2^k(v^k)&=\frac{I(M_2;Y_e^k|M_1,V^k=v^k)}{\sum_{i=1}^k (1-p_i)}.
\end{align}
Note that
\begin{align}
    \sum_{i=1}^k (1-p_i) S_1^k&=I(M_1;Y_e^k|M_2)
    \\&= I(M_1;Y_e^k,V^k|M_2)
    \\&= I(M_1;Y_e^k|M_2,V^k)
    \\&= \sum_{i=1}^k (1-p_i) \sum_{v^k} p(v^k) nS_1(v^k)
\end{align}
where we have used the fact that $V^k$ is a function of $Y_e^k$ and independent of $(M_1,M_2)$. Similarly, $S_2^k=\sum_{v^k} p(v^k) S_2(v^k)$. We may write
\begin{align}
    \sum_{i=1}^k (1-p_i)S_1^k(v^k)&=I(M_1;(Y_i:v_i=0,i\in[k])|M_2),\\
    \sum_{i=1}^k (1-p_i)S_2^k(v^k)&=I(M_2;(Y_i:v_i=0,i\in[k])|M_1).
\end{align}
Thus
\be
    (S_1^k(v^k),S_2^k(v^k))\in \frac{\sum_{i=1}^k (1-v_i)}{\sum_{i=1}^k (1-p_i)} \calC_{\text{TW}}.
\ee
By the convexity of $\calC_{\text{TW}}$,
\begin{align}
    (S_1^k,S_2^k)
    &=\sum_{v^k} p(v^k) (S_1^k(v^k),S_2^k(v^k))
    \\&\in \sum_{v^k} p(v^k)\frac{\sum_{i=1}^k (1-v_i)}{\sum_{i=1}^k (1-p_i)} \calC_{\text{TW}}
    \\&= \calC_{\text{TW}}
\end{align}
where we have used the fact that $\bbE[V_i]=p_i$.

For our MAC code, for any $k\in[0:n]$,
\begin{align}
    &n(R_1+R_2)=H(M_1,M_2)
    \\&\le I(M_1,M_2;Y_d^n)+n\eps_n
    \\&\le I(M_1,M_2;Y_d^n,Y_e^k)+n\eps_n
    \\&= I(M_1,M_2;Y_e^k)+\sum_{i=1}^n I(M_1,M_2;Y_{d,i}|Y_e^k, Y_d^{i-1})+n\eps_n\label{three_terms}
\end{align}
where $\eps_n$ is the Fano's inequality term, which goes to $0$ as $n\to\infty$. For the first term in \eqref{three_terms},
\begin{align}
    &I(M_1,M_2;Y_e^k)
    \\&=I(M_1;Y_e^k)+I(M_2;Y_e^k|M_1)
    \\&\le I(M_1;Y_e^k|M_2)+I(M_2;Y_e^k|M_1)\label{first_term1}
    \\&=\sum_{i=1}^k (1-p_i) (S_1^k+S_2^k)\label{first_term2}
    \end{align}
where \eqref{first_term1} follows because $M_1$ and $M_2$ are independent, and \eqref{first_term2} follows from the definitions of $S_1^k,S_2^k$.

For the second term in \eqref{three_terms}, for $i\le k$, we may write
\begin{align}
    &I(M_1,M_2;Y_{d,i}|Y_e^k, Y_d^{i-1})
    \\&\le I(M_1,M_2,X_{1,i},X_{2,i},Y_d^{i-1},Y_{e,i}^k;Y_{d,i}|Y_e^{i-1},V_i)
    \\&=p_i I(M_1,M_2,X_{1,i},X_{2,i},Y_d^{i-1},Y_{e,i}^k;Y_{i}|Y_e^{i-1},V_i=1)
    \\&=p_i I(X_{1,i},X_{2,i};Y_{i}|Y_e^{i-1})\label{markov1}
    \\&=p_i I(X_{1,i},X_{2,i};Y_{i}|U_i)
\end{align}
where \eqref{markov1} follows from the fact that, given $V_i=1$,
\be
(M_1,M_2,Y_d^{i-1},Y_{e}^k)\to (X_{1,i},X_{2,i})\to Y_{i}
\ee
is a Markov chain. For $i>k$, we have
\begin{align}
    &I(M_1,M_2;Y_{d,i}|Y_e^k, Y_d^{i-1})
    \\&\le I(M_1,M_2,X_{1,i},X_{2,i},Y_d^{i-1},Y_e^k;Y_{d,i}|V_i)
    \\&= p_iI(M_1,M_2,X_{1,i},X_{2,i},Y_d^{i-1},Y_e^k;Y_{i}|V_i=1)
    \\&= p_i I(X_{1,i},X_{2,i};Y_i)
\end{align}
where the last step follows from the same Markov chain as above. Putting this together, for each $k\in[0:n]$,
\begin{multline}
n(R_1+R_2)\le \sum_{i=1}^k\big[(1-p_i)(S_1^k+S_2^k)+ p_i I(X_{1,i},X_{2,i};Y_i|U_i)\big]
\\+\sum_{i=k+1}^n p_i I(X_{1,i},X_{2,i};Y_i)+n\eps_n.
\end{multline}
We also have
\begin{align}
    nR_1&=H(M_1|M_2)
    \\&\le I(M_1;Y_d^n|M_2)+n\eps_n
    \\&\le I(M_1;Y_d^n,Y_e^n|M_2)+n\eps_n
    \\&= I(M_1;Y_e^n|M_2)+I(M_1;Y_d^n|M_2,Y_e^n)+n\eps_n
    \\&= \sum_{i=1}^n\big[(1-p_i)S_1^n+ I(M_1;Y_{d,i}|M_2,Y_e^n,Y_d^{i-1})\big]+n\eps_n
    \\&= \sum_{i=1}^n\big[(1-p_i)S_1^n+p_i(H(Y_{i}|M_2,Y_e^n,Y_d^{i-1},X_{2,i})\nonumber
    \\&\qquad-H(Y_{d,i}|M_1,M_2,Y_e^n,Y_d^{i-1},X_{1,i},X_{2,i}))\big]+n\eps_n
    \\&\le \sum_{i=1}^n \big[(1-p_i)S_1^n+p_i(H(Y_{d,i}|U_i,X_{2,i})\nonumber
    \\&\qquad-H(Y_{d,i}|X_{1,i},X_{2,i},U_i))\big]+n\eps_n
    \\&= \sum_{i=1}^n\big[(1-p_i)S_1^n+ p_i I(X_{1,i};Y_i|U_i,X_{2,i})\big]+n\eps_n
\end{align}
where we have used the fact that $X_{2,i}$ is a function of $M_2$ and $Y_e^{i-1}$, as well as the same Markov chain as above. Similarly,
\be
nR_2\le \sum_{i=1}^n\big[(1-p_i)S_2^n+ p_i I(X_{2,i};Y_i|U_i,X_{1,i})\big]+n\eps_n.
\ee
Taking a limit as $n\to\infty$ concludes the proof.

\fi

\end{document}